\documentclass[prl,superscriptaddress,showkeys,showpacs,twocolumn]{revtex4-1}
\usepackage{graphicx}
\usepackage{latexsym}
\usepackage{amsmath}
\begin {document}
\title {Statistical mechanics model of angiogenic tumor growth}
\author{Ant\'onio  Luis Ferreira}
\affiliation{Departamento de  Fisica and I3N, Universidade de Aveiro, 3810-193 Aveiro, Portugal}
\author{Dorota Lipowska}
\affiliation{Faculty of Modern Languages and Literature, Adam Mickiewicz University, Pozna\'{n}, Poland}
\author{Adam Lipowski}
\affiliation{Faculty of Physics, Adam Mickiewicz University, Pozna\'{n}, Poland}
\begin {abstract}
We examine a lattice model of tumor growth where survival of tumor cells depends on the supplied nutrients. When such a supply is random, the extinction of tumors belongs to the directed percolation universality class. However, when the supply is correlated with distribution of tumor cells, which as we suggest might mimick the angiogenic growth, the extinction shows different, and most likely novel critical behaviour.  Such a correlation affects also the morphology of the growing tumors and  drastically raise tumor survival probability.
\end{abstract}

\maketitle
Due to series of mutations of  genes responsible for proliferation, some cells might initiate the  abnormal and uncontrolled growth process, commonly named as tumor \cite{nowak}.
In the first stage of the process, called avascular growth, there is no blood supply and the growth is limited by the amount of oxygen and nutrients that the tumor receives through its surface. The growing tumor initiates, however, a number of accompanying processes in the host environment. In particular, large demand for nutrients stimulates tumor cells to produce angiogenic factors that regulate the formation and growth of new blood vessels in the region. This stage is called angiogenesis and it intermediates between avascular and vascular growth. The vascular growth is the third stage, which begins when the blood vessels have reached the tumor.  In this stage the tumor receives a vast amount of nutrients and can grow much larger than it was possible during the avascular growth. Moreover, the vasculature might be used to spread tumor cells throughout the body of the host, which very often leads to its death. 

Tumor growth  is a very complex process and to fully understand its nature, one has to resort to computational techniques, which would supplement biological and medical approaches~\cite{byrne,bearer-oden,paulsonn,gabetta,bellomo}.  Various models were used to describe tumor growth \cite{tracqui}. Initially, they were continuous models formulated in terms of partial differential equations and studied mainly from mathematical point of view \cite{araujo}. More recently, inclusion of biomechanical details and coupling of tumor growth with the vascularization process shifted modeling toward more physical approach~\cite{preziosi}.

An alternative to differential-equation approach is based on discrete lattice models such as, for example, nonequilibrium $Q$-state Potts models. With such a modeling, one can implement several aspects of cell dynamics, which are difficult to treat simultaneously using continuous modeling, as e.g.\ multiplication, competition, aging, death, mutations and even adhesion or chemotaxis \cite{alber}. Some models of this kind were successfully tested against clinical data of certain forms of cancer \cite{torquato,jiang}. One can hope that further development will result in lattice models taking into account some other important, but so far neglected, factors like heterogeneity, immune cells or the role of chemoattractants.

Discrete lattice modeling allows us to describe tumor growth problem using tools developed in statistical mechanics for studying complex systems. A distinctive feature of these systems is spontaneous emergence of certain properties,  which cannot be traced to the character of individual parts. It is believed that life, consciousness, or functioning of ant colonies   are examples of such emergent phenomena \cite{yam}. There is an increasing evidence that cancer can also be considered as an emergent property, and thus, developing statistical-mechanics approaches seems to be very promising \cite{schwab,anderson,mansury}. In particular, one can hope that such bottom-up modeling will help us to explain tumor growth  in terms of cell parameters, which might contribute to its better prediction and control.

The ultimate goal, namely builiding realistic and testable against real data models, will most likely require very complex, multi-scale  models, which will be difficult  to understand without extensive computer simulations. To develop some intuitive understanding of tumor growth, it is thus desirable to examine some simpler models, which hopefully contain important ingredients of the process. In the present Letter, we examine  a simple lattice growth model where tumor cells survival and breeding depend on the supplied nutrients. Mimicking angiogenetic processes, we assume that the supply of nutrients is positively correlated with the distribution of tumor cells. It turns out that such a correlation substantially changes the statistical mechanics behaviour of the model. In particular, the extinction of tumor does not belong to the expected directed-percolation universality class  and the critical exponent~$\beta$ describing the order parameter most likely gets the classical value $\beta=1$, even in the $d=1$ version of the model. Such a correlation affects also the morphology of growing tumors and  drastically raises the tumor-survival probability.

In our model, each site of a $d$-dimensional lattice either is occupied by tumor, nutrient, both tumor and nutrient, or is empty. At a rate $p$, nutrients are supplied to a chosen site of the lattice, provided that the site is not occupied already by a nutrient. The roulette-wheel selection~\cite {liproulette} is used to choose the site for such a supply and the corresponding weight~$w$ depends on whether the site is occupied by tumor ($w=1+\Delta$) or not ($w=1$). The parameter $\Delta>0$ takes into account angiogenic effects of increased nutrients supply due to formation of new blood vessels in the vicinity of tumor cells.
At a rate $1-p$, a tumor cell on a randomly chosen site is updated. The tumor cell survives provided that there is a nutrient on this site, otherwise it dies. The surviving tumor cell consumes the nutrient and attempts to breed provided that there is a site without a tumor cell among its nearest neighbours. Closely related models but without any preference in nutrient supply ($\Delta=0$) were already studied~\cite{wendykier}.

To examine our model we used Monte Carlo simulations~\cite{preparation}. For various~$p$ and~$\Delta$, we measured the steady-state densities of tumor cells~$x_t$ and of nutrients~$x_n$. Simulations started from a random initial configuration and the model relaxed until a steady state was reached.
We examined lattices of various sizes~$N$ to ensure that the obtained results are $N$-independent. We also measured the time dependence of the tumor cell density~$x_t(t)$, with the unit of time corresponding to~$N$ update attempts. The density $x_t(t)$ for each $t$ is an average over independent runs. At the critical point, the density~$x_t(t)$ is expected to have a power-law decay $x_t(t)\sim t^{-\delta}$, where~$\delta$ is a characteristic exponent~\cite{hinrichsen}.

\begin{figure}
\includegraphics[width=9cm]{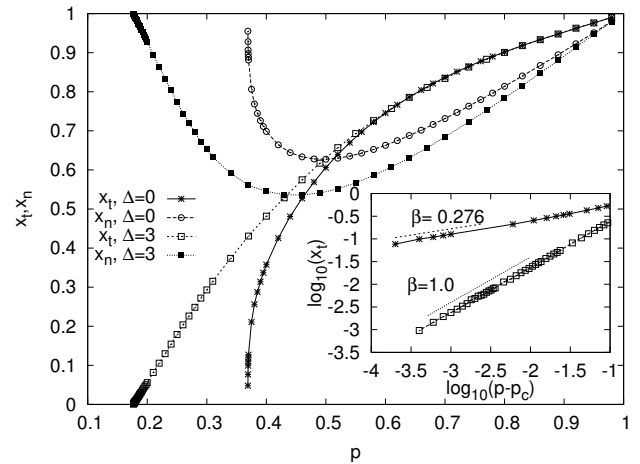}  \vspace{-0.7cm} 
\caption{Steady-state  densities of nutrients ($x_n$) and of tumor cells ($x_t$) as a function of $p$ calculated for $d=1$ and $\Delta=0$ and 3. The inset shows the behaviour of $x_t$ close to the critical point (log-log scale). While for $\Delta=0$ the directed-percolation scaling is seen, much different behaviour appears for $\Delta=3$. 
\label{steadyd1}}
\end{figure}

\begin{figure}
\includegraphics[width=9cm]{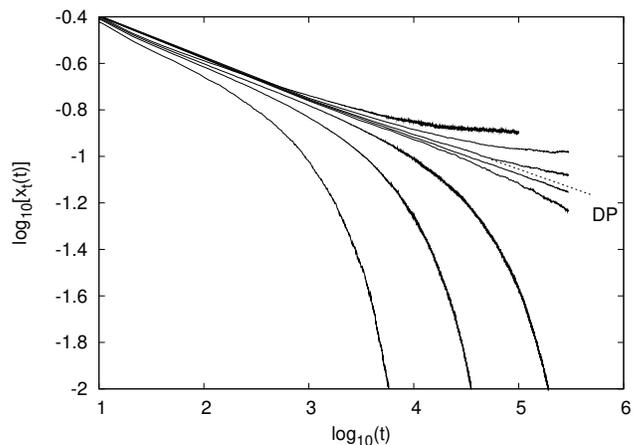}  \vspace{-1cm} 
\caption{Time dependence of the density of tumor cells $x_t(t)$ (log-log scale) for $d=1,\ \Delta=0$ and (from top) $p=0.37$, 0.3695, 0.3692, 0.36905, 0.3689  0.368, 0.366, and 0.36. The dotted straight line has the slope corresponding to the directed percolation value $\delta=0.159$.
\label{timedelta0}}
\end{figure}

First, we describe results for the $d=1$ version of the model. Simulations show that, for sufficiently large~$p$, the model remains in an active phase with $x_t>0$, which terminates at a critical point~$p_c$ depending on~$\Delta$. For $p<p_c$, the steady state of the model  is an absorbing state $x_t=0$ and $x_n=1$ (tumor cells die out of the lack of nutrients).  At $p=p_c$, the model undergoes the phase transition from an active into absorbing phase and we expect that $x_t$ is the corresponding order parameter. As it is already known, models with a single absorbing state are expected to belong to the so-called directed-percolation~(DP) universality class~\cite{hinrichsen}.  In this universality class and for $d=1$, the critical exponent $\delta_{DP}=0.159(1)$ and the decay of the order parameter upon approaching the critical point is described by the exponent $\beta_{DP}=0.276(1)$. 
The calculated steady-state values of $x_t$ and $x_n$ for $\Delta=0$ and 3 are presented in Fig.\ref{steadyd1}.
For $\Delta=0$, we estimate $p_c=0.3691(2)$ and the exponent $\beta$ is very close to the DP value (inset in Fig.\ref{steadyd1}). The decay of $x_t(t)$ at the critical point (Fig.\ref{timedelta0}) confirms the DP universality class in this case.

\begin{figure}
\includegraphics[width=9cm]{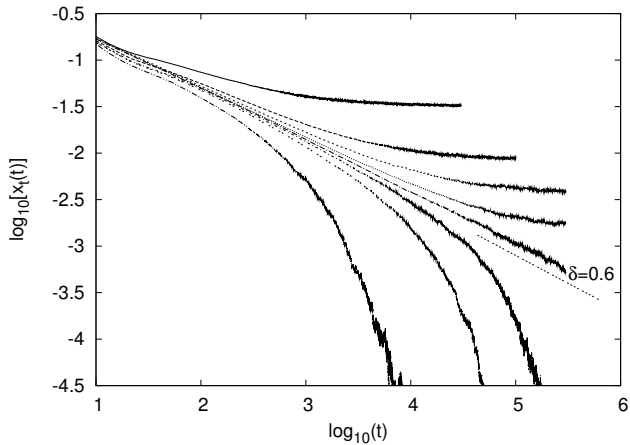}  \vspace{-1cm} 
\caption{Time dependence of the density of tumor cells $x_t(t)$ (log-log scale) for $d=1,\ \Delta=3$ and (from top) $p=0.19$, 0.18, 0.178, 0.177, 0.1765, 0.176, 0.175, 0.17. The dotted straight line has the slope corresponding to $\delta=0.6$.
\label{timedelta3}}
\end{figure}

Much different behaviour is seen for $\Delta=3$. In this case, we estimate $p_c=0.1765(2)$ and the exponent $\beta = 1.0(1)$ (inset in Fig.\ref{steadyd1}). The same estimate of $p_c$ is obtained from the behaviour of $x_t(t)$ (Fig. \ref{timedelta3}). However, at $p=p_c$ the decay seems to be described by the exponent $\delta = 0.60(5)$, which is again much different than the~DP value~0.159.  In calculations for $\Delta=0 \textrm{ and } 3$, and close to critical points, systems as large as $N=5{\rm x}10^5$ were used and simulation and relaxation times were of the order of  $10^6$. Further away from critical points, less extensive simulations were required.

\begin{figure}
\includegraphics[width=9cm]{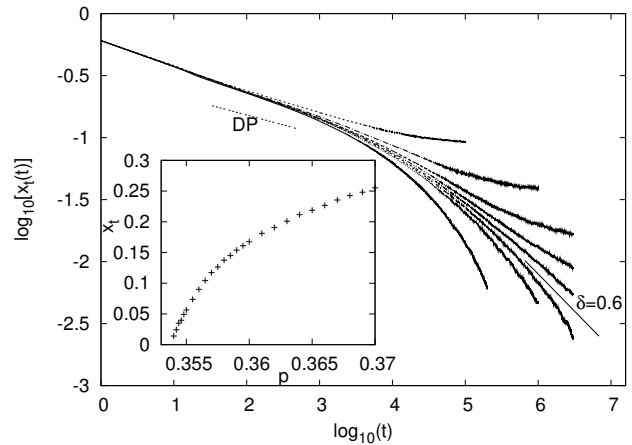}  \vspace{-1cm} 
\caption{Time dependence of the density of tumor cells $x_t(t)$ (log-log scale) for $d=1,\ \Delta=0.1$ and (from top) $p=0.356$, 0.3545, 0.354, 0.3538, 0.3537, 0.3536, 0.3535, and 0.353. The dotted straight line has the slope corresponding to $\delta=0.6$. Inset shows the steady-state density of tumor cells $x_t$ as a func\-tion of $p$. In the vicinity of the critical point, $x_t$  seems to decay linearly ($\beta=1$). 
\label{timedelta01}}
\end{figure}

For $\Delta=0.1$  on a shorter time scale ($t\sim 10^2$), one can notice a slower, DP-like decay of $x_t(t)$,  that, however, turns to much faster decay on a longer time scale (Fig.\ref{timedelta01}). Such a behaviour indicates the proximity of the~DP behaviour that  occurs at $\Delta=0$. However, critical exponents $\beta$ and $\delta$ for $\Delta=0.1$ are nearly the same as for $\Delta=3$. We do not present numerical results, but simulations show that the same estimations of critical exponents are obtained for $\Delta=5$ and~10. 
Hence, our results suggest that the critical exponents $\beta = 1.0(1)$ and $\delta = 0.60(5)$ are universal most likely for any $\Delta>0$.    

Let us notice that for increasing~$\Delta$, the critical rate~$p_c$ decreases and the tumor-free phase shrinks. Thus, angiogenic factors, that in our model correspond to positive correlation between nutrient  supply and tumor cell distribution ($\Delta>0$) make starvation of tumors to death more difficult.  Such a behaviour of our model is very plausible since angiogenesis was "invented" by tumors just for this particular reason. For very large~$\Delta$, the critical rate~$p_c$ seems to vanish, which means that tumors can survive even under a very small rate of the nutrient supply. Although the presented model is too simple to describe detailed complexity of real tumors, in our opinion, it qualitatively correctly captures the role of angiogenic factors.
 
Despite possessing a single absorbing state, our model most likely for any $\Delta>0$ does not belong to the directed-percolation universality class.  The exponent $\beta$ seems to take the classical, mean-field  value $\beta=1$ that for a $d=1$ model is certainly a puzzling result.  The exponent $\delta=0.60(5)$ is different than the mean-field value $\delta=1$~\cite{hinrichsen} and than the directed-percolation value (0.159). Possibly novel critical behaviour of our model is confirmed with calculation of other critical exponents,  but detailed presentation of our results will be given elsewhere \cite{preparation,different}. 
\begin{figure}
\vspace{-3.cm} 
\includegraphics[width=9cm]{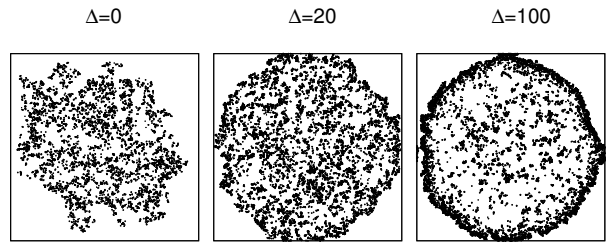}  \vspace{-1cm} 
\caption{Different morphologies of $d=2$ tumors growing from a single tumor cell surrounded by empty lattice sites. Calculations were made on 500x500 lattices and $p$ was chosen such that the steady-state density in all three cases was approximately equal ($x_t\approx 0.05$).
\label{config}}
\end{figure}

We also examined the $d=2$ version of our model. Obtained results~\cite{preparation} suggest that the values of critical exponents are very close to those in the $d=1$ version. The $d=2$ version of our model is certainly more realistic in the context of tumor-growth modeling. We examined their growth starting from a configuration containing a single tumor cell and all other sites empty. Simulations were performed for various~$p$ and~$\Delta$  but such that the steady-state density of tumor cells $x_t$ was approximately the same and equal to $0.05$. Numerical simulations show that $\Delta$ strongly affects the shape and dynamics of growing tumors (Fig.\ref{config}). While for small $\Delta$ their shape is highly irregular, for large $\Delta$ nearly circular shape with a well defined boundary is seen. Apparently, in the former case the growth is very much affected by stochastic fluctuations while in the latter their role  is diminished.  

\begin{figure}
\includegraphics[width=7.0cm]{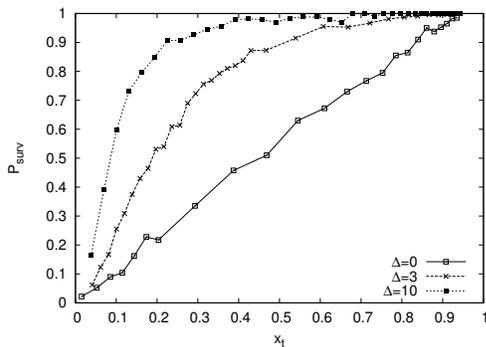}  \vspace{-0.7cm} 
\caption{Tumor survival probability $P_{\rm surv}$ as a function of steady-state density $x_t$. Initial configuration contained a single tumor cell surrounded by empty lattice sites.       Calculations were made on 200x200 lattice and evolution of tumors was monitored until $t=10^4$. Calculations were made for several values of $p$ and the corresponding steady-state density of tumors $x_t$ was estimated from steady-state simulations, similarly to the $d=1$ case shown in Fig. \ref{steadyd1}.
\label{surv}}
\end{figure}

Different morphologies of growing tumors suggest that their other characteristics will also strongly depend on $\Delta$. We measured the tumor survival probability starting from the same configurations containing a single tumor cell and monitored whether tumor cells survived until a given (large) simulation time. The simulations were performed for several values of~$p$ and~$\Delta$ and the survival probability $P_{\rm surv}$ is plotted as a function of the steady-state density of tumors~$x_t$, that was estimated independently with standard steady-state simulations (Fig.~\ref{surv}). Let us notice that some general arguments developed for models with absorbing states suggest that in the active phase, both~$P_{\rm surv}$ and the steady-state order parameter should scale with the same critical exponent~$\beta$~\cite{hinrichsen}. Thus,  one expects an approximately linear dependence  $P_{\rm surv}\sim x_t$ for sufficiently small~$x_t$.
Numerical results  confirm that $P_{\rm surv}$ increases linearly with~$x_t$ but they also show a strong dependence on~$\Delta$.  Indeed, for the same~$x_t$, the tumor survival probability for  $\Delta=3$ and 10  is much larger than in the $\Delta=0$ case. This is yet another indication of the importance of angiogenic effects on the tumor growth.

In conclusion, we examined a lattice model of tumor growth where the positive correlation of  nutrient supply with tumor cells distribution mimics angiogenic  factors. Obtained results show that such a correlation shifts the location of a tumor extinction and makes their starvation to death more difficult. Moreover, it  changes the critical behaviour of the model, even though as a model with a single absorbing state it should belong to the directed percolation universality class. Surprisingly, even in the $d=1$ version the exponent $\beta$ seems to take the classical, mean-field value $\beta=1$. However, the exponent $\delta$ describing the time decay of the order parameter at criticality takes a non-classical value 0.60(5) and most likely the model represent a novel critical behaviour. Field-theory methods were applied to various models with absorbing states \cite {hammal} and one can hope that also in this case they could provide valuable insight.  In the $d=2$ version such a correlation affects the morphology of growing tumors and substantially raises their survival probability. From the statistical mechanics perspective it would be desirable to examine the behaviour of our model for $-1<\Delta<0$ as well as in the limit $\Delta\rightarrow\infty$, where the model simplifies and nutrient supply is forbidden to occur on empty sites.


\begin{thebibliography}{}
\bibitem{different} We checked  that decay of the order parameter  at criticality is described by nearly the same exponent ($\delta=0.60(5)$) that does not depend on the class of initial configurations. In particular, we made simulations starting from configurations containing various concentrations of tumor cells and nutrients.
\end{thebibliography}
\end {document}